\documentclass
[preprint,pra,showpacs,showkeys,titlepage,balancelastpage]{revtex4}%
\usepackage{amsfonts}
\usepackage{amsmath}
\usepackage{amssymb}
\usepackage{graphicx}%
\begin{document}
\title[Gedanken experiments]{Gedanken experiments with Casimir forces, \\vacuum energy, and gravity }
\author{G. Jordan Maclay}
\email{jordanmaclay@quantumfields.com}
\affiliation{Quantum Fields LLC, Richland Center WI 53581}
\volumeyear{2010}

\begin{abstract}
Gedanken experiments are used to explore properties of quantum vacuum energy
that are currently challenging to explore experimentally. \ A
constant\ lateral Casimir force is predicted to exist between two overlapping
finite parallel plates at 0 K, otherwise it would be possible to extract an
arbitrary amount of energy from the quantum vacuum. \ A rigid unpowered object
cannot be accelerated by the quantum vacuum because of the translational
symmetry of space. By considering systems in which vacuum energy and other
forms of energy are exchanged, \ we demonstrate that a change $\Delta E$ in
vacuum energy, whether positive or negative with respect to the free field,
corresponds to an equivalent inertial mass and equivalent gravitational mass
$\Delta M=\Delta E/c^{2}.$ We consider the possibility of a gravitational
shield, and show that, if it exists, the energy to operate it would have to
cancel the net energy extracted from the gravitational field, otherwise we
could extract an arbitrary amount of energy from the field.

\end{abstract}
\startpage{101}
\endpage{10}
\maketitle

\section{Introduction}

The Casimir force per unit area between two infinite, parallel, perfectly
conducting plates is given by $F(a)=-K/a^{4}$ where $K=\pi^{2}\hslash c/240$
$=1.3\times10^{-27}Nm^{2}$ , and $\hslash$ is Plank's constant, $c$ is the
speed of light in vacuum. \ This force arises because the plates change the
vacuum energy density between the plates from the free field energy density.
\ Although the force was predicted by Casimir in 1948, it is so small, even at
distances of several tenths of a micrometer that a quantitative measurement
was not made until 1998, when an Atomic Force Microscope (AFM) was used to
measure the force between a sphere and a plate to an accuracy of
1\%\cite{mohideen}. The challenge of securing parallelism between plates with
submicron separations has limited the accuracy of force measurements between
two plates to about 15\% \cite{bressi}.

We are interested in considering several aspects of vacuum energy and Casimir
forces, including the inertial mass associated with vacuum energy, the
interaction of vacuum energy and gravity, and the possibilities of utilizing
vacuum energy for propulsion or other purposes. \ There are three conceptual
types of grav-inertial mass: inertial mass that resists acceleration, active
gravitational mass that generates a gravity field around it, and passive
gravitational mass that reacts to a gravitational field. \ These terms arise
in the Parameterized Post Newtonian expressions for gravitational\ energy and
force and are discussed by Will \cite{will}. \ These terms all can conceivably
be positive, negative, imaginary, complex, position dependent, anisotropic.
\ Some of them can be conceivably identical. \ Newtonian mechanics and General
Relativity assume that inertial mass, active gravitational mass, and passive
gravitational mass are identical, positive and isotropic, and no experiments
to date have contradicted these assumptions. \ The equivalence principle
assumes that inertial mass and passive gravitational mass are identical, and
independent of material, and the measurements to date have not contradicted
this assumption \cite{will}. \ The notion of inertial mass arises in special
relativity as the Lorentz invariant norm $p^{\mu}p_{\mu}$ of the
energy-momentum four-vector $(E,$ $pc)$, namely$\ \ \ p^{\mu}p_{\mu}%
\ =E^{2}-p^{2}c^{2}=m^{2}c^{4}$. \ In the rest frame of the particle, the
momentum $p$ is zero, so the energy is $mc^{2}$ and $m$ is called the rest
mass of the particle. \ The existence of inertial mass can be seen as a
consequence of the four dimensional symmetry of space-time.

Experimentalists measuring Casimir forces have looked for a modification to
the usual force of gravity at short distances as proposed by Fischbach
\cite{fischbach}, but to date no such modifications have been
found\cite{decca}.

There is no generally accepted theory of inertial mass\cite{will}. \ A recent
proposal based on interactions with the vacuum field is
controversial\cite{puthoff}, but nevertheless the vacuum field does seem to be
a factor. \ In conventional quantum electrodynamics, radiative shifts arise
from the interactions of a particle with the zero point fluctuations of the
vacuum electromagnetic field. \ The real part of the shift is a mass shift.
\ The vacuum field can be interpreted as jostling an idealized point particle,
giving it kinetic energy and an equivalent mass\cite{weiskopf}. \ The
amplitude of the motion is too small to be observed directly, but changes in
the vacuum field can result in measurable changes in the mass. \ Thus if the
vacuum field is altered from the free field, mass shifts occur. \ For example,
a spinless particle near a surface experiences a different vacuum field than
if very far from the surface, which results in a shift in the effective
mass\cite{bartonpp}.\ If there were a generally accepted theory of inertial
mass, then it might be somehow "different" in definition and perhaps behavior
than gravitational mass. \ Even so, it is highly probable that active
gravitational mass and passive gravitational mass are identical and positive
for all ordinary matter we know. \ Very puzzling dynamics can occur if they
are not equal.

Gravity is generated by the local energy-momentum tensor source term in the
Einstein gravity equation, which is a function of mass, energy, linear
momentum, angular momentum, stress, charge, spin, etc. \ Some contributions,
like the "gravitational twist" that angular momentum makes, are gravity fields
that are distinguishable from the radial Newtonian gravity field of the rest
mass. \ If the ground state electromagnetic energy in the quantum vacuum were
treated like any other form of energy, it would be expected to produce a
corresponding gravitational field, and changes in the energy would be expected
to produce changes in the gravitational field. \ There are, of course, some
severe problems reconciling quantum field theory with general relativity:
\ the current theory would regard the infinite quantum vacuum energy density
as a gravitational source term, the effect of which would be to rip apart the
universe\cite{weinberg}. \ Although this inconsistency between two widely
accepted theories has not been resolved, the general consensus is that only
changes in vacuum energy act as a source of a gravitational field.

Calculations of vacuum stresses for a variety of geometric shapes, such as
spheres, cylinders, rectangular parallelepipeds, and wedges are reviewed
in\cite{kimbalbook}\cite{mostbook}\cite{milonni} .There are complications and
problems with the computation of vacuum energies of objects and surfaces,
especially divergences in the energy which arise from curvature in the
surfaces\cite{deutschandcandelas}, such as right angles, or from ideal
boundary conditions, such as perfect conductors\cite{dewitt1}\cite{maclaypra}.
\ The material properties, such as the frequency dependent dielectric constant
and plasma frequency of the metal and the surface roughness affect the vacuum
forces. \ In addition, in the usual calculations only a spatial average of the
force for a given area for the ground state of the quantum vacuum field is
computed, and material properties, such as binding energies, are ignored, a
procedure which Barton has questioned\cite{bartonshapes}. \ Difficulties is
defining vacuum energies for spheres and other shapes have been discussed by
Graham et al\cite{graham}. \ We will consider the vacuum force for a very
simple case of two finite plates sliding over each other with a fixed separation.

There are other less dramatic difficulties in distinguishing the energy and
mass of particles in gravitational fields, for example the difficulties of
including the effect of gravity on the mass of a particle, as in a system of
two particles with a gravitational potential energy between the masses that
results in a gravitational binding energy which changes the effective particle
mass\cite{fulling}. \ There are even difficulties with the notion of a
particle in curved spaces\cite{birrell}.

From quantum field theory, it appears that only the changes in vacuum energy
from the free field values are meaningful experimentally, since we can compute
these changes only, not the absolute value of the energy, which generally is
divergent. \ For example, the Casimir energy\ density for a parallel plate
geometry is actually the difference between the free field vacuum energy
density and the vacuum energy density between the plates. \ Similarly the Lamb
shift in the hydrogen atom can be computed as the real part of the shift in
the atomic energy level that arises when the atom is placed in the quantum
vacuum. \ It can also be computed as the corresponding shift in the energy of
the quantum vacuum that occurs when hydrogen atoms are put in the quantum
vacuum \cite{milonni}. \ The latter shift arises from the change in the
frequencies of the vacuum field due to the change in the index of refraction
from the hydrogen atoms. \ It makes sense to propose that a shift in vacuum
energy corresponds to a shift in the inertia of an object
\cite{reynaudquantvacfluc}.\ \ To verify this hypothesis, Reynaud calculated
the effective inertial mass of two parallel mirrors that are coupled by the
vacuum field, and found that this mass included the shift in the vacuum energy
between the two plates\cite{reynaudmotional}. \ Changes in vacuum energy
density result in changes in the effective mass.

When we consider vacuum energy, there are three questions that we will
address: 1. \ Is a change in inertia of a system associated with a change in
the vacuum energy of the system? \ 2. Is a gravitational field generated by
the change in vacuum energy (equivalent active gravitational mass)? \ 3. If an
external gravitational field is present, is there a change in the
gravitational energy of the system that is associated with the change of
vacuum energy (equivalent passive gravitational mass)? \ We will consider
these questions and several other questions using a variety of gedanken
experiments\cite{gedanken}. \ 

The benefit of this approach is that we can assume a vastly simpler landscape
than the formal approach using quantum electrodynamics and general and special
relativity, omitting details of specific systems and the explicit
consideration of divergences. \ This simplification can highlight the role of
basic concepts and clarify and generalize the essential physics. \ On the
other hand, what is lost are the details, for example the presence of
divergences, the explicit consideration of the Hamiltonian of the vacuum
field. \ It is therefore very reassuring that for the cases in which formal
calculations\ have been attempted, there is agreement with our gedanken
results. \ For the parallel plates geometry, Milton et al computed that
changes in vacuum energy correspond to changes in inertial mass and couple to
gravity in the same way as conventional forms of energy \cite{miltonvacen}%
\cite{miltoncasenfall}.

The gedanken experiment was a powerful tool in the hands of Einstein. \ He
described a gedanken experiment in which he said he "demonstrated that the
mass of a body is a measure of its energy content; if the energy changes by
$E$, the mass changes by $E/c^{2}.$"\ \ Probably his most famous gedanken
experiment was the EPR description of entangled states. \ Unfortunately he
never considered vacuum energy.

Another benefit of gedanken experiments is that we consider systems that may
be very difficult or impossible to investigate experimentally. \ Many of the
effects related to vacuum energy are extremely small. \ Nevertheless they are
of fundamental interest. For example, the Schornhorst effect\cite{schorn}%
\cite{bartonschorn}, which predicts that light moving transversely between
parallel plates propagates faster than $c,$ but the effect is so small it
cannot be directly measured \cite{milonnischorn}. \ Negative vacuum energy
acts like a negative mass. \ Calloni et al \cite{calloni} have considered the
repulsive gravitational force due to the negative vacuum energy in a stack of
$10^{6}$parallel plate capacitors, and found that it is slightly beyond the
current capability for measurement using the most sophisticated gravitational
wave detection technology.

In addition to the fundamental understanding of vacuum energy and gravitation,
we are interested in the potential role of vacuum energy in space
travel\cite{reynaudspace}\cite{puthoffvac}\cite{frontiers}. \ Hence several
space motivated gedanken experiments are included. \ Science fiction author
Arthur C. Clark, who proposed geosynchronous communications satellites in
1945, described a"quantum ramjet drive" in 1985 in "Songs of Distant Earth",
and observed in his Acknowledgement, "If vacuum fluctuations can be harnessed
for propulsion by anyone besides science-fiction writers, the purely
engineering problems of interstellar flight would be solved"\cite{clark}.
\ Recently Australian writer Ken Ingle described a quantum vacuum powered
engine\cite{ingle}. \ There have been numerous papers on space warps and
drives that often presuppose the ability to generate material with negative
mass, or generate macroscopic gravitational fields by manipulation of vacuum
energy\cite{morris}. \ One proposal is to employ the Casimir effect to reduce
the vacuum energy density below the free field value, but this effect is very
small, as Calloni computed\cite{calloni}. \ Unfortunately, these interesting
ideas are well beyond any technology that we can foresee. \ Without some
breakthrough, such as a new boundary condition on the vacuum that causes much
greater energy shifts, interstellar exploration appears impossible.

For simplicity in this paper we will only consider the free quantum vacuum
field at zero absolute temperature. \ In the gedanken experiments, we will
assume that for ordinary matter the active, passive, and inertial masses are
identical, and that for ordinary energy $E$, such as chemical or mechanical,
the contribution to inertial mass is given by $E=mc^{2}$.

The first gedanken experiment demonstrates that there must be a lateral
Casimir force acting when one finite flat plate slides over another perfectly
parallel finite flat plate. \ If this lateral Casimir force were not present,
it would be possible to extract an arbitrary amount of energy from the quantum
vacuum. \ (In Casimir research, the same phrase, "lateral force", has been
used to describe vacuum forces between corrugated surfaces\cite{chenlateral}.)

The second gedanken experiment shows that a rigid unpowered object cannot be
accelerated in the quantum vacuum unless some of the mass of the object is
being converted to energy directly, as in radioactive decay. \ If an unpowered
object could be accelerated by the quantum vacuum, it would, in principle, be
possible to extract an unlimited amount of energy from the vacuum, and we
would have a continuous acceleration for a spacecraft with no expenditure of energy.

The third gedanken experiment demonstrates that we need to associate an
inertial mass $\Delta m$ with changes $\Delta E$ in vacuum energy according to
$\Delta E=\Delta mc^{2}$. \ The fourth, fifth, and sixth experiments show that
changes in vacuum energy correspond to equivalent active and passive
gravitational masses. \ Changes in vacuum energy couple to the gravitational
field like other forms of energy, otherwise one could continuously extract
energy from a gravitational field.

The seventh gedanken experiment is motivated by science fiction writers
proposing gravitational shields. \ We consider the existence of a box which
would insulate the mass inside from the effects of an external gravitational
force. \ This device would lead to the paradox of being able to extract energy
continuously from the gravitational field, unless the energy required to open
and close the box just canceled the extracted energy.

\section{Gedanken Experiments}

\subsection{Gedanken Experiment One: \ Existence of Finite Flat Plate Lateral
Casimir Force}

In this gedanken experiment we consider the energy balance when we move
parallel conducting plates through a cycle of both lateral and transverse
motions\cite{maclaymanual}. \ Initially we have two perfectly conducting,
completely overlapping $(x=L)$, square, parallel plates, a distance $L$ on
each side, that are a distance $a$ apart, with $a<<L$. If we allow the upper
plate to approach the lower (fixed) plate quasistatically, then the attractive
Casimir force $F_{C}(a)=-KL^{2}/a^{4}$ does positive mechanical work during
this reversible thermodynamic transformation.%
\begin{figure}
[ptb]
\begin{center}
\includegraphics[
trim=0.000000in 0.059586in 3.065142in 0.109696in,
height=1.5134in,
width=2.6783in
]%
{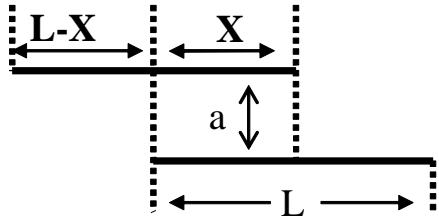}%
\caption{Two square L x L parallel perfectly conducting plates with an
overlapping area L x X, separated by a distance $a$. Only lateral motion is
permitted.}%
\label{Fig 1}%
\end{center}
\end{figure}
We are neglecting all edge effects by assuming that the force is proportional
to the exact area of overlap. \ During the transformation, the vacuum energy
$E_{C}(a)=-KL^{2}/3a^{3}$ between the plates will be reduced, conserving the
total energy in the system. If the separation decreases from $a_{i}$ to
$a_{f}$ , then the energy balance is%
\begin{equation}
E_{C}(a_{f})-E_{C}(a_{i})=-\int_{a_{f}}^{a_{i}}F_{C}(a)da
\end{equation}

If we then separate the plates quasistatically, letting $a$ increase from
$a_{f}$ to $a_{i}$, we do work on the system to restore it to its initial
state. Over the entire cycle, no net work is done, and there is no net change
in the vacuum energy.

Consider an alternative cycle that has been proposed in order to extract
energy from the vacuum fluctuations: After the plates have reached the point
of minimum separation, slowly slide the upper plate laterally until it no
longer is opposite the lower plate $(x=0)$, which eliminates the normal
Casimir force, then raise the upper plate to its original height, and slowly
slide it laterally over the lower fixed plate $(x=L)$. \ Finally we allow the
plates to come together as before, extracting energy from the vacuum
fluctuations and doing mechanical work. If no energy was expended in moving
the plates laterally, then this cycle would indeed result in net positive work
equal to the energy extracted from the vacuum. \ Although no one has yet
computed in detail the lateral forces between offset finite parallel plates,
it is highly probable that such forces exist, and that no net extraction of
energy occurs for this cycle.

We can verify this by a simple approximate calculation. We neglect Casimir
energy \textquotedblleft fringing fields,\textquotedblright\ and assume that
the energy density differs from the free field density only in the region in
which the two square $(L\times L)$ plates overlap a amount $x$, where $0<x<L$
(see Fig. 1). \ Then we can compute the lateral force $F_{L2}$ between the two
plates using the conservation of energy (principal of virtual work):%
\begin{equation}
F_{L2}(x)=-d[-KLx/3a^{3}]/dx=KL/3a^{3} \label{lat}%
\end{equation}

where $a$ ($a<<L)$ is the perpendicular distance between the plates. \ This
constant (independent of $x$) attractive lateral force tends to increase $x$
or pull the plates towards each other so they have the maximum amount of
overlap and minimum vacuum energy. \ In fact, the positive work done to move
one plate laterally a distance $L$ exactly cancels the work extracted from the
vacuum fields in moving the plates from a large separation to a distance $a$
apart, so there is no net change in total energy (mechanical plus field) in
the complete cycle, as expected. \ 

The normal Casimir force between these $L\times L$ plates when they are
directly opposite, with complete overlapping ($x=L$), is $L/3a$ times larger
than the constant lateral force given by Eq. \ref{lat}. \ Experimental
verification of lateral forces on flat plates is challenging but may be
possible. \ This force could be determined by a measurement in the shift of
the average position
$<$%
x%
$>$
of a vibrating surface opposite the edge of a fixed surface. \ One
experimental arrangement would be to have a vertical cantilever with a flat
horizontal plate on the top. \ This vertical cantilever could be vibrated
horizontally in the x-direction and its deflection measured. \ Then a second
cantilever with a ball on the end, horizontal, would be mounted on a xyz
piezostage allowing it to move in the y-direction (up and down), as well in
the x-direction (and in and out). \ The ball would be brought down so its
center would be directly above the edge of the plate on the vertical
cantilever. \ The ball has an approximately fixed x coordinate, and the y
coordinate must be close enough so the Casimir force is measurable. \ Then the
plate will experience a lateral force. \ If the vertical cantilever is
vibrating, then the lateral force from passing over one edge of the plate will
result in a net force that will shift the mean location of the plate as the
ball moves closer to the plate. \ The vertical Casimir force will also be changing.

Numerous investigators have considered from a theoretical perspective the
situation of two infinite plates at zero Kelvin sliding at constant velocity
over one another. \ Some researchers have concluded a quantum friction is
present and some have not. \ These efforts were recently reviewed by Philbin
and Leonhart \cite{philbin}, who computed that there is no quantum friction
for this situation, although there is a higher order modification of the
transverse Casimir force due to the velocity. \ Not all researchers agree with
their conclusions. \ As they mention, the situation for finite plates is quite
different, which our gedanken experiment confirms. \ In the gedanken
experiment we assumed the motion was slow, and neglected velocity corrections
to the transverse Casimir force.

\subsection{Gedanken Experiment Two: \ No Quantum Vacuum Sails}

This gedanken experiment shows that no rigid, unpowered object can experience
a net acceleration in the quantum vacuum, unless its mass is being directly
converted to energy, as in radioactive decay. \ Imagine an object in the free
field isotropic vacuum, distant from any other objects, whose geometry is
fixed. \ The object might be composed of various materials, with various
dielectric coefficients, in thermal equilibrium, and with a fixed arbitrary
shape. \ Assume the object does not contain any power supply, mechanical or
electric. \ It is generally quite difficult to explicitly compute the vacuum
stress on such an object, however, if the object did experience a net
acceleration in the vacuum, then one could, in principle, use the movement of
the object to operated a machine, and extract an arbitrary amount of energy
from the vacuum.

First consider a plane surface because a variety of sail concepts have been
proposed \cite{millis} . \ We can view the vacuum as a source of radiation
pressure from virtual photons\cite{milonni}. \ The challenge is to design
surfaces that alter the symmetry of the free vacuum and produce a net force.
\ Consider for example, a sail made of two different materials on opposite
sides, that absorb electromagnetic radiation differently. \ Can we expect a
net force on the sail? \ A simple classical analysis as shown in Fig 1
suggests the answer to this question\cite{maclayaiaa}.%
\begin{figure}
[ptb]
\begin{center}
\includegraphics[
trim=0.007819in 0.008955in 2.314495in 0.010348in,
height=1.6034in,
width=2.981in
]%
{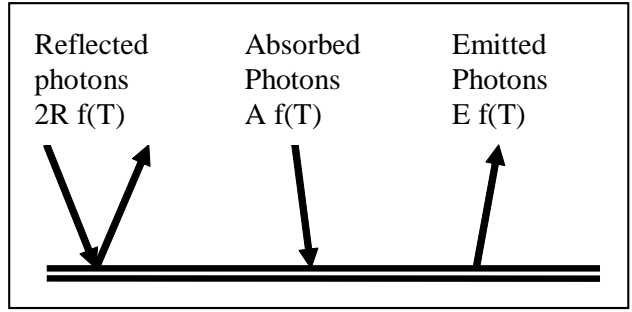}%
\caption{Schematic of the momentum transfer from electromagnetic radiation to
a sail made of different materials on the top and bottom.}%
\label{FIg 2}%
\end{center}
\end{figure}
\ 

For a given frequency, assume the radiation energy density is proportional to
$cf(\omega,T)$, so the net momentum transfer $\Delta P_{\omega}$\ to the top
surface is%
\begin{equation}
\Delta P_{\omega}=A_{\omega}f(\omega,T)+E_{\omega}f(\omega,T)+2R_{\omega
}f(\omega,T)
\end{equation}
where $A_{\omega}$ is the absorptivity, $E_{\omega}$ is the emissivity,
$R_{\omega}$ the reflectivity, and $T$ the temperature. \ \ For a body in
thermodynamic equilibrium, $A_{\omega}=E_{\omega}$, and by definition,
$1=A_{\omega}+R_{\omega}$. \ Using these restriction, it follows that $\Delta
P_{\omega}=2f(\omega,T),.$which is independent of the material
properties.\ \ Therefore the force on the opposite side of the sail just
cancels this force, and there is no net acceleration. This conclusion holds at
every frequency.

We assumed the temperature of the sail is the same on both sides because of
the intimate contact. \ If the radiation spectrum corresponds to that at zero
temperature, the zero point field, then both sides of the sail would be at
zero Kelvin. \ On the other hand, there would be a net force if one side of
the sail was pointed at a source of photons (such as our sun) causing a
different radiation density on one side of the sail than the other. \ If one
made a powered sail in which a temperature gradient was maintained across the
sail, a net force could occur, and it would be a function of the energy
required to maintain the temperature difference.

For the most general rigid, unpowered object, consider that the Hamiltonian
$H$ of the object depends on the various internal coordinates $(q_{i},p_{i})$
corresponding to the objects geometry. \ We assume because of the
translational invariance of space that the energy of the object does not
depend on the location of the center of mass, for which the corresponding
operator is $\mathbf{Q,}$ nor does it depend explicitly on time. \ Then it
follows:
\begin{equation}
\lbrack H(q_{i},p_{i},\mathbf{P}_{i}),\mathbf{P}_{j}]=i\hbar\frac{\partial
H}{\partial Q_{j}}=0
\end{equation}
where $P_{j}$ =-i$\hbar\partial/\partial Q_{j}$ is the operator for the $j$
component of the center of mass momentum. \ The Hamiltonian might depend on
the center of mass momentum $\mathbf{P}$. \ Since the Hamiltonian is also the
generator of the translations in time, it follows that%
\begin{equation}
\lbrack H(q_{i},p_{i},\mathbf{P}_{i}),\mathbf{P}]=i\hbar\frac{\partial
\mathbf{P}}{\partial t}=0
\end{equation}
\ and the momentum of the center of mass is conserved.

There are two possibilities. 1.\ The inertial mass remains constant, so the
center of mass velocity must also be constant and there is no acceleration;
\ 2. The inertial mass and velocity both vary but in a way that conserves the
center of mass momentum. \ However, such a variation would not be consistent
with a constant kinetic energy of the center of mass. \ This would imply that
there must be a compensating change in another form of energy within the
system. \ In effect, the mass energy is being converted to kinetic energy.
\ This might be due to the decay of an excited or radioactive atom emitting
particles or radiation. \ However, if we do not have decay or some similar
process\ converting mass to energy, then we conclude that there is no net
acceleration of our object. \ 

This gedanken experiment shows, for example, that neither a passive air foil,
nor a rigid open cavity (box with no top on it) can accelerate by itself in
the vacuum.

Although this result might seem obvious or trivial, there are some assumptions
and subtleties. \ We have not explicitly included a Hamiltonian for the vacuum
fields and therefore we have not explicitly considered the role of curvature
and singularities in the energy momentum tensor for the vacuum field. \ Since
no real photons are generated by a curved surface, one would not expect this
to alter our conclusion. \ On the other hand, if nuclei with $Z\alpha>1$ were
present, then real photons could be produced from the quantum vacuum, so we
are excluding this possibility. \ Another subtlety is that near a surface, the
vacuum force shows small fluctuations, and therefore there will be small
variations in the net force, that tend to accelerate the surface,
statistically in random directions\cite{bartonfluc}. \ From dimensional
considerations the root mean square value of the fluctuations in the vacuum
force scale as $\hbar/c^{3}T^{8},$ where $T$ is the time interval between
measurements. \ The effect is negligible except for extraordinarily short
times. \ Calculation shows that the effect may add slightly to the
fluctuations based on the time-energy uncertainty relation. \ In the same
spirit, Rueda has suggested that very high energy particles observed in space
may derive their kinetic energy from a long term acceleration due to the
stochastic vacuum field\cite{rueda}.

The assumption that the device is not powered arises because of the
possibility that if the device were powered, then it could be accelerated in
the vacuum. \ Consider, for example, an anharmonically vibrating plate which
causes the emission of photons by the adiabatic Casimir effect\cite{maclayfop}%
. \ The radiative reaction will result in a small net acceleration of the
plate. \ In this example of a vibrating plate, the Hamiltonian must include
the radiation field of the photons which is correlated with the moving center
of mass.

One wonders if there are other unrecognized subtleties in the seemingly
innocuous assumptions that may modify this gedanken experiment. \ For example,
perhaps there may be presently unknown distortions or excitations in the
vacuum field that do not correspond to the emission of photons but that
nevertheless carry momentum and energy, like a form of dark energy-momentum.

\subsection{Gedanken Experiment Three: Vacuum Energy Contributes to Inertial
Mass}

From general relativity, various conventional forms of energy $E$ are
considered to contribute to the inertial mass as given by the equation
$E=mc^{2}.$ \ This gedanken experiment is designed to show that a change in
vacuum energy also gives a corresponding change in inertial mass. \ Imagine an
apparatus, the details of which will be described later, that is contained
within a small sphere with uniform, rigid, insulating walls so the system is
closed. \ The sphere is small enough so that it can serve as an inertial
frame. \ We assume that no heat, thermal radiation, energy or mass pass
through the walls, which do not vibrate. \ The sphere is embedded in the
quantum vacuum, far from any other objects or gravitational forces. \ The
Casimir force on the sphere will be inward, and therefore will not tend to
accelerate the center of mass of the sphere. \ As mentioned before, to a
higher order in $\hbar$ there are\ stochastic forces that tend to accelerate
the sphere in random directions, as in Brownian motion\cite{bartoncavityelec}.
\ As a practical matter these can be neglected over the duration of the
experiment. \ We assume that the sphere is not moving initially, or is in
uniform non-relativistic motion. \ For simplicity we consider only the vacuum
field at zero Kelvin; for fields at higher temperatures there are, as Einstein
proved, additional forces present \cite{maclayejop}. \ We assume vacuum
fluctuations are present within the sphere.

Within the sphere is a system consisting of parallel Casimir plates, a battery
which powers a motor which can change the spacing between the plates, and a
timer switch that control when the motor turns on and off (Figure 3). \ When
the plates are moved closer together quasistatically, the vacuum energy
between them decreases (becomes more negative), and work is done on the motor,
charging the battery. \ Assuming no dissipative forces, the total energy in
the sphere is conserved, and the decrease in the vacuum energy between the
plates equals the increase of the chemical energy in the ideal battery. \ This
process is reversible. \ (The same results would be obtained if we used a
coiled spring with a mechanical linkage instead of a motor and battery.) \ The
question is: does this transformation of energy alter the motion of the
sphere?%
\begin{figure}
[ptb]
\begin{center}
\includegraphics[
trim=0.214728in -0.042894in 1.321451in 0.459934in,
height=1.9726in,
width=2.7605in
]%
{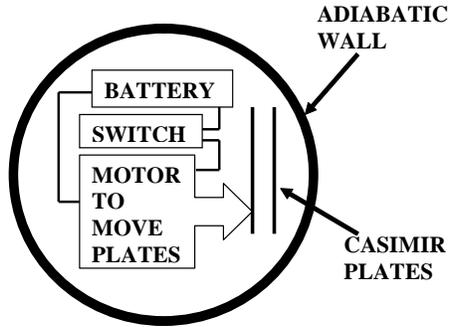}%
\caption{Schematic showing apparatus described in gedanken experiment 3,
consisting of\ a motor/generator to alter the spacing between two parallel
plates, and a battery to power the motor, a timer switch, all surrounded by a
wall impervious to thermal energy and matter.}%
\label{Fig 3}%
\end{center}
\end{figure}
As in Gedanken Experiment 2, no external force has acted on the closed system
while the spacing between the plates was changed, so the momentum and energy
must remain constant. \ As before, there are two possibilities:\ 1. the
velocity and the inertial mass are constant, or\ 2. the inertial mass
decreases and the velocity increases. \ 

If the velocity and inertial mass both changed in a manner that keep the
momentum constant, then the kinetic energy would have to change. \ However, an
increase in kinetic energy would violate conservation of energy, therefore a
sphere in the quantum vacuum would not accelerate. \ In other words, it is not
possible for a single moving body to conserve momentum and kinetic energy
unless the velocity is constant. \ Since\ the center of mass velocity does not
change as the plate separation is decreased, if follows from the conservation
of momentum that the inertial mass is constant.

We conclude that vacuum energy contributes to inertial mass in the same way as
the chemical energy in the battery.

The same result would follow for other forms of energy. \ For example, assume
we used a spring to drive the motion of the plates instead of a motor/battery
combination. \ The vacuum energy would contribute to the inertial mass just as
the potential energy in the spring contributed. \ Even if dissipative forces
were present, and vacuum energy was degraded to heat, the total energy would
remain constant as would the inertial mass. \ All forms of energy, including
quantum vacuum energy, contribute to the inertial mass according the general relativity.

The total energy in the sphere is conserved, although the form of the energy
may change. \ Within the container, there may be quasistatic, adiabatic
transformations, in which one type of energy increases, while another type
decreases, so the total remains constant. \ This suggests that one can
associate an increase in effective inertial mass with one element and a
decrease of effective inertial mass with another element. \ The increase and
decrease are with respect to the original state. \ This suggests that it might
be possible to make components that have negative inertial mass. \ Such
objects would tend to rise in a uniform gravitational field. \ Indeed negative
vacuum energy in the stack of parallel plate capacitors considered
theoretically by Calloni et al resulted in a force in a gravitational field
that was in the opposite direction from that experienced by normal positive
matter, but the positive force due to the mass of the silicon wafers, was much
larger. \ Could one make an object that floated in a gravitational
field?\ \ Since the objects we know of are composed of material with a
positive mass, it is not clear that one can make an object whose overall mass
is negative. \ The total energy in a Casimir parallel plate arrangement, even
with plates an atom thick, remains positive \cite{visser}.

To get a sense of the magnitude of forces involved consider two square
parallel plate $L$ long separated by a distance $a$\ in a gravitational field
$g.$ \ Then the total attractive Casimir force between the plates is $F_{C}=$
$-L^{2}K/a^{4}$. \ To lowest order in $g$, the gravitational force $F_{g}$ on
the parallel plates due to the negative vacuum energy $E_{C}=-aL^{2}%
(K/3a^{4})$ is \cite{calloni}
\begin{equation}
F_{g}=\frac{-gE}{c^{2}}=\frac{-gL^{2}}{c^{2}}\frac{K}{3a^{3}}%
\end{equation}
The ratio of this gravitational force to the Casimir force is%
\begin{equation}
\frac{F_{g}}{F_{C}}=\frac{ag}{3c^{2}}%
\end{equation}
For a typical plate separation $a\sim100nm$, the effective mass of the Casimir
energy $E_{C}$ is minus that of about 450 protons per cm$^{2}.$ \ The ratio of
the gravitational force to the Casimir force is is $1.1\times10^{-23}.$
\ Although this number is incredibly small, current gravity wave detectors can
monitor the position of a test mass to one part in $10^{21}$, so at some point
it may be possible to conduct an experiment\cite{nature}.

\subsection{Gedanken Experiment Four (falling sheres):\ Vacuum Energy
Contributes to Passive Gravitational Mass}

This gedanken experiment is designed to show that a shift in vacuum energy
gives a corresponding passive gravitational mass. \ We imagine two spheres as
described in the previous experiment, with Casimir plates, batteries, and
motors, falling near each other in the quantum vacuum in a weak gravitational
field. \ We assume that the chemical energy of the battery gives a passive
gravitational mass that couples to gravity in the normal way. \ On the other
hand, we assume that vacuum energy does not couple to gravity. \ As one sphere
is falling, the plate spacing remains fixed, while in the other sphere the
motor alters the spacing between the plates, converting chemical energy from
the battery into changes in the quantum vacuum energy between the parallel
plates. \ By our assumption, the acceleration of the second sphere will
increase or decrease as energy is transferred from the battery to the Casimir plates.

In this system, the kinetic energy of both spheres is increasing with time.
\ A change in the acceleration of one sphere relative to the other sphere
would require an additional external force, which is not present. \ Hence, the
acceleration must not change with changes in vacuum energy. \ This shows that
changes in the quantum vacuum energy give rise to corresponding changes in the
passive gravitational mass and inertial mass. \ Vacuum energy couples to
gravity the same way any other form of energy is expected to couple to gravity.

\subsection{Gedanken Experiment Five (explicit calculation in the
gravitational field of a mass): Vacuum Energy Contributes to Passive
Gravitational Mass}

In this experiment, we consider the coupling of vacuum energy to a
gravitational field of a mass M. \ In the gedanken experiment, we will
determine the quantitative consequences if vacuum energy does not couple to
the gravitational field. \ Consider the same apparatus used in previous
gedanken experiments consisting of a battery with chemical energy $U_{B}$,
motor/generator and Casimir plates with vacuum energy $U_{C}$. \ The apparatus
is initially at a distance $R_{1}$ from a gravitational mass $M$. \ The
initial gravitational potential energy of the chemical energy of the battery
is
\begin{equation}
U_{i}=-GMU_{B}/(c^{2}R_{1}).
\end{equation}
\ By assumption, there is no potential energy corresponding to the vacuum
energy $U_{C}.$ \ \ Assume we have a device, such as a motor/generator and
rope, that can lower the apparatus from $R_{1}$ to a distance $R_{2}$ from the
mass $M$. \ When this is done, the lowering device will have net positive work
done on it, and the potential energy of the apparatus will decrease, but the
sum of both will remain constant since energy is conserved. \ If we raise the
apparatus back to $R_{1}$, then this net positive energy of the lowering
device is used up, and there is no net change in energy in the system since
the gravitational field is conservative. \ We assume there is no friction or
other dissipative force, and we neglect the mass of the rope in the calculation.

Now imagine lowering the apparatus again from $R_{1}$ to $R_{2}$. \ Once the
apparatus is at $R_{2}$, the the gravitational potential energy of the
chemical energy in the battery is $-GMU_{B}/(c^{2}R_{2}).$ \ \ The amount of
work done by the lift device to lower the apparatus equals the change in
potential energy%
\begin{equation}
W_{d}=GMU_{B}(R_{2}^{-1}-R_{1}^{-1})/c^{2}%
\end{equation}
Assume the battery now turns on, which sends energy $E$ to the motor which
increases the separation of the Casimir plates, which increases the vacuum
energy to $U_{C}+E$. \ Conversely the battery energy is reduced by the same
amount to $U_{B}-E$ , so the battery is lighter. \ Its potential energy at
$R_{2}$ is reduced to $-GM(U_{B}-E)/(c^{2}R_{2}).$ \ We assume that the energy
in the quantum vacuum does \textbf{not} couple to gravity, so there is no
increase in gravitational potential energy corresponding to the change in
vacuum energy $E$.

Imagine now raising the apparatus from $R_{2}$ to $R_{1}.$ \ Less work will be
done to raise the apparatus to $R_{1}$ than before since the battery is
lighter. \ At $R_{1}$ the potential energy of the battery is
\begin{equation}
U_{f}=-GM(U_{B}-E)/(c^{2}R_{1}).
\end{equation}
$\ $The amount of work done by the lift device is
\begin{equation}
W_{u}=-GM(U_{B}-E)(R_{2}^{-1}-R_{1}^{-1})/c^{2}%
\end{equation}
\ and the energy of the Casimir plates remains $U_{c}+E.$

Once the apparatus is at $R_{1},$ \ we imagine extracting vacuum energy $E$
from the Casimir plates so the vacuum energy is now $U_{C}$ and charging the
battery to its original energy state $U_{B}.$ \ This conversion will result in
an additional gravitational potential energy of
\begin{equation}
U_{E}=-GME/(c^{2}R_{1}).
\end{equation}
\ The system has been returned to its original state, but there is a net
increase in energy of the system. \ The net change in energy of the system
equals the total energy of the final state minus the energy of the initial
state:
\begin{equation}
\Delta E=W_{d}+W_{u}+U_{f}+U_{E}-U_{i.}=GME(R_{2}^{-1}-R_{1}^{-1})/c^{2}.
\end{equation}

There is a net increase in energy of the system but no change in the state of
the system. \ This is a clear violation of the conservation of energy. \ Hence
our assumption is not valid and we must conclude that vacuum energy couples to
the gravitational field like any conventional form of energy.

\subsection{Gedanken Experiment Seven: Vacuum Energy Contributes to Active
Gravitational Mass}

In order to show that vacuum energy contributes to active gravitational mass,
we consider a variation on the experimental arrangement in the preceding
gedanken experiment. \ We have a fixed apparatus, consisting of the motor,
battery, and Casimir plates, and we assume its equivalent active gravitational
mass is $M.$ \ Assume we have a test mass $m$ separated from the sphere by a
distance $R_{1}$. \ We then move the mass $m$ until it is a distance $R_{2}$
from the apparatus. \ The change in potential energy is $-GMm(R_{2}^{-1}%
-R_{1}^{-1}).$ \ Then we increase the plate separation using energy $E$ from
the battery, which reduces the active gravitational mass of the battery by
$E/c^{2}$ and the active gravitation mass of the apparatus to $M-E/c^{2}.$
\ We assume the change in vacuum energy does not change the\ equivalent active
gravitational mass. \ We now move the mass $m$ back to its original location,
doing an amount of work $-G(M-E/c^{2})m(R_{1}^{-1}-R_{2}^{-1})$. \ We then use
the battery to operate the motor and move the plates towards each other until
they are at their original separation. \ An energy $E$ is extracted from the
vacuum and is used to charge the battery to its original energy state. \ This
causes a shift in the potential energy of of the mass m equal to
$-Gm(E/c^{2})(R_{1}^{-1}).$ \ The system has been returned to its original
state and there is net increase in energy equal to $G(E/c^{2})(R_{2}^{-1})$.
\ This violates the conservation of energy. \ Hence our assumption that vacuum
energy does not contribute to active gravitational mass is not true.

\subsection{Gedanken Experiment Eight:\ No Free Energy with Gravity Shields}

This experiment explores gravitational shields, the stuff of science fiction.
\ A few experiments have been done, for example, with rotating superconductors
to determine if there is any evidence of gravitational shielding, with null
results \cite{frontiersgrav}. \ If such shielding devices were possible, how
would they operate? \ What would be their limitations?

We consider a box with special walls that totally shield the interior of the
box from any external gravitational field. \ The box has a door which can be
opened and closed to insert a mass $M$. \ We assume that the inertial mass of
$M$ is not affected by the box. \ Assume the gravitational potential energy of
the mass $M$ is $U_{1}$ when we insert it into the box and close the door.
\ (Closing the door can be understood as a euphemism for "turning on" the
gravity shield for whatever is inside the box.) \ For simplicity, we assume
that gravity does not exert any force on the box. \ Now imagine moving the box
to a different location. \ Since there is no external force of gravity on the
box and the box is stationary at the beginning and end of the movement, no net
work is done. \ Imagine we now open the door, and remove the mass. \ At this
new location the gravitational potential energy of the mass $M$ is $U_{2}$.
\ By the conservation of energy, the change in potential energy $U_{2}-$
$U_{1}$ should equal the work done on the system. \ By our assumptions, no net
work was done to move the box, so we conclude to not violate the conservation
of energy we must do an amount of work $U_{2}-$ $U_{1}$ to operate the door of
the box. \ In general, the amount of work necessary to operate the door will
equal the difference in energy between the mass $M$ at its final location and
its initial location. \ Negative work $-$ $U_{1}$ is done to close the door
(it could close by itself), and positive work $U_{2}$ must be done to open the
door (or turn off the gravity shield).

For example, imagine we put a space capsule into the box. \ We then accelerate
the box to begin an interstellar trip. \ No energy is used to overcome
gravitational fields, only to overcome inertia, reducing fuel needs by several
orders of magnitude. \ At the end of the trip, on some distant planet, the
energy to open the box will simply be the change in potential energy.
\ Conceivably, opening and closing the door might be done en route, near
gravitational sources, as part of the navigational technology. \ 

If we were to put the space capsule into the box on earth, and shut the door,
then the earths radial gravitational acceleration would suddenly disappear,
and the mass would accelerate tangentially to the earths surface at about 1000
miles/hour, an interesting way to use the earths rotational velocity to launch
a space capsule. \ If the accelerating mass pushed against the wall of the
gravitational box, and accelerated the box tangent to the earth's surface,
then in a simple geometric model (neglecting air resistance and the effect of
gravity on the box itself) the box would be about 100 miles above the surface
of the earth after one hour.

This gedanken experiment may be based on a material that is impossible to
make. \ Using the analogy from electrostatics, shielding depends on the
existence of positive and negative charge, whose effects can cancel each
other. \ An atom of antimatter could indeed cancel the gravitational energy of
an atom of matter, but they do not coexist in any known form, so the existence
of a gravity shield might actually violate physical laws.

This observation that gravity shielding may be impossible brings to mind a
recent proposal regarding the theoretical expressions of Lifshitz which are
used to model Casimir forces for real materials \cite{kimbalbook}%
,\cite{mostbook} \ A suggestion was made that the Lifshitz theory needed to be
modified to account for screening effects and diffusion currents\cite{dalvit}.
\ The Lifshitz theory of Casimir forces assumes thermal equilibrium. \ On the
other hand, diffusion currents and screening effects occur when thermal
equilibrium is not present. \ It appears that including these effects violates
thermal equilibrium, and hence is not consistent with the basic Lifshitz
formulation \cite{mostepanenko}. \ This illustrates the subtleties that may
lie in seemingly innocuous assumptions about screening. \ Would a box that
shields against vacuum fluctuations be fundamentally impossible?

\section{Conclusion}

Gedanken experiments are used to explore properties of vacuum energy that are
currently challenging or impossible to explore experimentally. \ A
constant\ lateral Casimir force is predicted between two overlapping finite
parallel plates, otherwise it would be possible to extract an arbitrary amount
of energy from the quantum vacuum. \ By considering systems in which vacuum
energy and other forms of energy are exchanged, \ we demonstrate that a change
$\Delta E$ in vacuum energy, whether positive or negative with respect to the
free field, corresponds to an equivalent inertial mass and gravitational mass
$\Delta M=\Delta E/c^{2}.$

The first gedanken experiment demonstrated that there is a constant, finite
lateral force at 0 K between two parallel, finite plates that overlap. \ The
force tries to maximize the amount of overlap. \ Other gedanken experiments
have shown that changes in vacuum energy formally couple to gravity like
ordinary forms of energy. \ Otherwise, it is possible to design gedanken
experiments in which an arbitrary amount of energy can be extracted from a
physical system without changing the state, violating our usual form of the
law of conservation of energy. \ Specifically, changes in vacuum energy
correspond to equivalent active and passive gravitational masses. \ Positive
shifts in vacuum energy act like ordinary matter; whereas negative shifts in
vacuum energy correspond to negative masses, which are repelled by the
gravitational force with ordinary matter. \ This unusual property of negative
vacuum energy makes it very interesting, since it might allow, in principal at
least, the formation of structures which have zero equivalent mass, and the
cancellation of gravitational forces. \ Unfortunately, in practice, the
methods used to generate the negative vacuum energy, for example, Casimir
plates, are so limited in the negative energy density they can produce, that
it does not appear possible, without some new approach, to make an actual
object that has net zero or a negative vacuum energy. \ Perhaps, in
astrophysical systems, other boundary conditions pertain, and larger negative
vacuum energies are possible.

Within the next decade, experiments may be done to verify some of the
conclusions drawn from the gedanken experiments, for example, the lateral
Casimir force. \ Extracting energy from the quantum vacuum is clearly possible
if there is a change in the state of the system. \ It is done when the spacing
between the Casimir plates is changed by the motor/battery combination in our
gedanken experiments. \ Experiments on the exchange of energy between the
quantum vacuum and ordinary physical systems will help us understand the role
of vacuum energy. \ It is possible that new methods, new boundary conditions,
will be found that can be used to extract large amount of energy from the
quantum vacuum. \ Cole has considered this possibility in an astrophysical
situation\cite{colethermo}.

\begin{acknowledgments}
We have had the good fortune to have had fruitful exchanges about vacuum
energy and gravitational interactions with the late Bryce DeWitt, the late
Robert Forward, Gabriel Barton, Peter Milonni, Dan Cole, Hal Puthoff, and An
T. Nguyen Le. \ We also thank the referee for helpful suggestions.
\end{acknowledgments}

\end{document}